\documentclass[a4paper,11pt]{article}
\pdfoutput=1

\usepackage{jheppub}

\usepackage{graphicx} % Required for inserting images
\usepackage{hyperref}
\usepackage[utf8]{inputenc}
\usepackage{amssymb}
\usepackage{multirow}
\usepackage{soul}

\usepackage{booktabs}
\usepackage{mathrsfs}
\usepackage{epsfig}
\usepackage{graphicx}%Include figure files
\usepackage{subfigure}
\usepackage{dcolumn}%Align table colums on decimal point
\usepackage{bm}%bold math
\usepackage{amsmath}
\usepackage{slashed}
\usepackage{multirow}
\usepackage{color}
\usepackage{dcolumn}
\usepackage{bm}
\usepackage{color}

\allowdisplaybreaks

 \newcommand{\lsim}{{\;\raise0.3ex\hbox{$<$\kern-0.75em\raise-1.1ex\hbox{$\sim$}}\;}}
\newcommand{\gsim}{{\;\raise0.3ex\hbox{$>$\kern-0.75em\raise-1.1ex\hbox{$\sim$}}\;}}
\newcommand{\beq}{\begin{equation}}
\newcommand{\eeq}{\end{equation}}
\newcommand{\bea}{\begin{eqnarray}}
\newcommand{\eea}{\end{eqnarray}}

\def\baa{\begin{array}}
\def\eaa{\end{array}}

\mathchardef\minus="002D

\preprint{ }

\title{Probing Type II Seesaw Leptogenesis Through Lepton Flavor Violation}

\author{Chengcheng Han$^{1,2}$,}
\emailAdd{hanchch@mail.sysu.edu.cn}

\author{Yijun Han$^1$,}
\emailAdd{shlghyj1998@126.com}

\author{Sihui Huang$^{1}$,}
\emailAdd{huangsh99@mail2.sysu.edu.cn}

\author{Zhanhong Lei$^{1}$}
\emailAdd{leizhh3@mail2.sysu.edu.cn}

\vspace{0.5cm}

\affiliation{$^1$School of Physics, Sun Yat-Sen University, Guangzhou 510275, P. R. China}
\affiliation{$^2$Asia Pacific Center for Theoretical Physics, Pohang 37673, Korea}

\vspace{0.5cm}

\abstract{Lepton flavor violation (LFV) offers a powerful probe of physics beyond the Standard Model, particularly in models addressing neutrino masses and the baryon asymmetry of the universe. In this study, we investigate LFV processes within the framework of type II seesaw leptogenesis, where the Standard Model is extended by an $SU(2)_L$ triplet Higgs field. We focus on key LFV processes including $\mu^+\to e^+\gamma$, $\mu^+ \to e^+e^-e^+$, and $\mu \rightarrow e$ conversion in nuclei, deriving stringent constraints on the parameter space from current experimental data. We scan the 3$\sigma$ range of neutrino oscillation parameters and identify the most conservative bounds consistent with existing measurements. Our results reveal that the MEG experiment currently provides the strongest constraints in the normal ordering (NO) scenario, while the SINDRUM experiment offers comparable sensitivity in the inverted ordering (IO) case. Future experiments, such as MEG II, Mu3e, Mu2e, and COMET, are predicted to significantly improve the sensitivity, testing larger regions of the parameter space. This work underscores the crucial role of LFV experiments in probing type II seesaw leptogenesis, providing an avenue to explore the connections between neutrino mass generation, baryogenesis, and inflation at experimentally accessible energy scales.}

\makeatletter
\def\@fpheader{\relax}
\makeatother

\date{2023.05.18}

\begin{document} 
\maketitle
\flushbottom
\newpage

\section{Introduction}

It is well-known that in the Standard Model the lepton flavor $L_e,~L_\mu,~L_\tau$ are conserved, arising as accidental symmetries. However, with the discovery of neutrino oscillations\,\cite{Cleveland:1998nv,Super-Kamiokande:1998kpq,SNO:2002tuh,SNO:2003bmh,KamLAND:2002uet,K2K:2006yov,DayaBay:2012fng}, it is realized that the lepton flavor is not a symmetry in nature. Although charged lepton flavor violation(CLFV) has not yet been observed, numerous neutrino mass models have provided such sources. Therefore, the CLFV experiments have become excellent probes of new physics, shedding light on the origin of neutrino masses\,\cite{Bernstein:2013hba,Lindner:2016bgg,Calibbi:2017uvl,Ardu:2022sbt,Frau:2024rzt}.

Among the various neutrino mass models, the seesaw models are the most popular\,\cite{Minkowski:1977sc,Yanagida:1979as,Gell-Mann:1979vob,Magg:1980ut,Cheng:1980qt,Lazarides:1980nt,Schechter:1980gr,Mohapatra:1980yp,Foot:1988aq}. Moreover, the lepton number violation and CP violation in seesaw models naturally lead to the mechanism called leptogenesis\,\cite{Fukugita:1986hr,Albright:2003xb}, which is an appealing candidate for the explanation of baryon asymmetry of our universe. However, leptogenesis through type I or type III seesaw generally involve high energy scales\,\cite{Giudice:2003jh}, thus are difficult to probe. On the other hand, it is known that type II seesaw mechanism can not lead to successful thermal leptogenesis\,\cite{Ma:1998dx,Hambye:2005tk}. Interestingly, recent works show that by considering Affleck-Dine mechanism and non-minimal coupling of the scalar fields to gravity, it may be possible to realize successful non-thermal leptogenesis in type II seesaw mechanism\,\cite{Barrie:2021mwi,Barrie:2022cub}. Thus one may explain the origin of neutrino mass, inflation and baryon asymmetry simultaneously in the framework of type II seesaw leptogenesis. 

An attractive feature of type II seesaw model is that the relevant scale is relatively low among the numerous new physic models. This means it can be tested by current and future upcoming experiments, both in the high energy frontier with collider experiments\,\cite{Melfo:2011nx,Du:2018eaw,Antusch:2018svb,BhupalDev:2018tox,Ashanujjaman:2021txz,Chun:2019hce,Mandal:2022ysp,Han:2023vme} and in the intensity frontier with charged lepton flavor violation(CLFV) experiments\,\cite{Chun:2003ej,Kakizaki:2003jk,Akeroyd:2009nu,Dinh:2012bp,Han:2021nod,Ferreira:2019qpf,Barrie:2022ake,Calibbi:2022wko,Ardu:2023yyw,Ardu:2024bua}. Since both classes of experiments have obtained null result, constraints can be placed on the parameter space of the model. In this paper our discussions would focus on the CLFV tests of type II seesaw leptogenesis, however, most of our methods and results are generic to any type II seesaw models. Analogous analyses have been conducted in \,\cite{Barrie:2022ake}, however, they only obtained the constraints at several benchmark points of the neutrino oscillation  parameters. In this paper, we scan the 3$\sigma$ range of the neutrino data
 %the whole parameter space with Monte Carlo method
and obtain the most conservative constraints on the parameters of type II seesaw leptogenesis. In addition, comparing the analysis on $\mu \rightarrow e$ conversion in the titanium(Ti) nuclei in \,\cite{Barrie:2022ake},  we also include the $\mu \rightarrow e$ conversion in aluminum(Al) nuclei. In this work, we mainly focus on the constraints of CLFV experiments from the muon sector, including the searching for $\mu^+\to e^+\gamma$, $\mu^+ \to e^+e^-e^+$ and $\mu^- N \to e^- N$ processes. Current best limits on these processes are provided by the MEG, SINDRUM and SINDRUM II experiments respectively\,\cite{Adam:2013vqa,SINDRUM:1987nra,SINDRUMII:1993gxf,SINDRUMII:2006dvw}. Moreover, several upcoming experiments will improve the sensitivity by up to four orders of magnitude\,\cite{MEGII:2018kmf,Mu3e:2020gyw,Mu2e:2014fns,COMET:2018auw,CGroup:2022tli}, which will greatly extend the parameter region we are interested in. 

This paper is organized as follows. In Sec.\,\ref{section2} we briefly review the framework of type II seesaw leptogenesis. In Sec.\,\ref{section3} we calculate the CLFV processes in the muon sector and obtain the current most stringent constraints on the parameter space of type II seesaw leptogenesis. Then, we analyses the sensitivity of future CLFV experiments on the model. Finally in Sec.\,\ref{conclusion} we give a summary and draw conclusions.

\section{Type II Seesaw Leptogenesis}
\label{section2}

\subsection{Type II Seesaw Mechanism}

In the type II seesaw mechanism, the SM is minimally extended by an $SU(2)_L$ triplet Higgs field $\Delta$ with hypercharge $Y=1$ and lepton number $L=-2$:
\begin{equation}
    \Delta=\left(
    \begin{array}{cc}
        \Delta^+/\sqrt{2} & \Delta^{++} \\
        \Delta^0 & -\Delta^+/\sqrt{2} \\
    \end{array} \right),
\end{equation}
and the SM Higgs is
\begin{equation}
    H=\left(\begin{array}{cc}
        \phi^+ \\
        \phi^0
    \end{array}\right)~.
\end{equation}
The Lagrangian of the type-II seesaw scenario is given by
\begin{equation} 
\label{LagrangiantypeII}
\mathcal{L}^{\text{II}}_{\text{seesaw}}=\mathcal{L}_{SM}+\text{Tr}\left[(D_{\mu}\Delta)^\dagger\left(D^{\mu}\Delta\right)\right]-V\left(H,\Delta\right)-\mathcal{L}_\text{Yukawa},
\end{equation}
where
\begin{equation}
\mathcal{L}_\text{Yukawa}=\mathcal{L}_\text{Yukawa}^\text{SM}-h_{ij}\bar{L}^c_i(i\sigma^2)\Delta L_j+h.c.~.
\end{equation}
The Higgs potential $V(H,\Delta)$ has the form,
\begin{eqnarray}
\label{potential}
V\left(H,\Delta\right)&=&-m_H^2H^\dagger H+\lambda_H\left(H^\dagger H\right)^2+m_\Delta^2 \text{Tr}\left(\Delta^\dagger\Delta\right)+\lambda_1\left(H^\dagger H\right)\text{Tr}\left(\Delta^\dagger\Delta\right)\nonumber\\
&&+\lambda_2\left(\text{Tr}\left(\Delta^\dagger\Delta\right)\right)^2+\lambda_3 \text{Tr}\left(\Delta^\dagger\Delta\right)^2+\lambda_4 H^\dagger\Delta\Delta^\dagger H\nonumber+\left[\mu(H^Ti\sigma^2\Delta^\dagger H) \right.\nonumber \\
   && \left. +\frac{\lambda_5}{M_P}(H^Ti\sigma^2\Delta^\dagger H)(H^\dagger H)+ \frac{\lambda_5^\prime}{M_P}(H^Ti\sigma^2\Delta^\dagger H)(\Delta^\dagger \Delta) +h.c.\right]~,
\end{eqnarray}
where the terms in the bracket violate the lepton number, which are necessary for leptogenesis during inflation. At low energy scales where we are interested in, only the $\mu-$term contributes. 

After electroweak symmetry breaking(EWSB), the neutral components of $H$ and $\Delta$ obtain nonzero vacuum expectation value(vev). In the limit $m_\Delta\gg v_{EW}$, which is consistent with the constraints from LHC\,\cite{ATLAS:2017xqs}, we get the tiny triplet vev:
\begin{equation}
\label{tripletHiggsvev}
v_\Delta\equiv\langle\Delta^0\rangle\simeq\frac{\mu v^2_{EW}}{2m^2_\Delta}
\end{equation}
where $v_{EW}\simeq174\text{\,GeV}$. Once the neutral component of the triplet Higgs $\Delta^0$ obtain its non-zero vev, a tiny neutrino mass is generated from the Yukawa term. The neutrino mass matrix can be computed as
\begin{equation}
\label{neutrinomass}
\left(m_\nu\right)_{ll^{\prime}}\equiv m_{ll^{\prime}}\simeq 2h_{ll^{\prime}} v_\Delta~.
\end{equation}
Thus the matrix of Yukawa couplings $h_{ll^{\prime}}$ is directly related to the PMNS neutrino mixing matrix $U_\text{PMNS}\equiv U$,
\begin{equation}
\label{Yukawacouplings}
    {h_{ll^{\prime}}}\equiv{\frac{1}{2v_\Delta}\left(U^* \text{diag}\left(m_1,m_2,m_3\right)U^\dagger\right)_{ll^{\prime}}}~.
\end{equation}
where $m_i$ are the masses of the neutrino mass eigenstates, and $U$ is the PMNS matrix which takes the form,
\begin{equation}
U=V\left(\theta_{12},\theta_{23},\theta_{13},\delta\right)Q\left(\alpha_{21},\alpha_{31}\right)~,
\end{equation}
where
\begin{equation}
    V=\left(
    \begin{array}{ccc}
    1 & 0 & 0 \\
    0 & c_{23} & s_{23} \\
    0 & -s_{23} & c_{23} \\
    \end{array}\right)
    \left(\begin{array}{ccc}
    c_{13} & 0 & s_{13}e^{-i\delta}\\
    0 & 1 & 0 \\
    -s_{13}e^{i\delta} & 0 & c_{13}\\
    \end{array}\right)
    \left(\begin{array}{ccc}
    c_{12} & s_{12} & 0 \\
    -s_{12} & c_{12} & 0 \\
    0 & 0 & 1 \\
    \end{array}\right), Q=\left(\begin{array}{ccc}
    1 & 0 & 0 \\
    0 & e^{i\alpha_{21}/2} & 0 \\
    0 & 0 & e^{i\alpha_{31}/2}
    \end{array}\right)~,
\end{equation}
and $c_{ij}=\text{cos}\theta_{ij}$, $s_{ij}=\text{sin}\theta_{ij}$, $\delta$ is the Dirac phase, and the matrix $Q$ contains the two Majorana phases, $\alpha_{21}$ and $\alpha_{31}$. The best fit values of these parameters obtained from latest neutrino oscillation experiments are given in Table.\ref{table}\,\cite{Esteban:2020cvm}.

\begin{table}
\renewcommand\arraystretch{1.2}
\tabcolsep=0.7cm
\begin{tabular}{|c|cc|}
\hline $\mathrm{BFP} \pm 1 \sigma$ & $\mathrm{NO}$ & $\mathrm{IO}$ \\
\hline $\sin ^{2} \theta_{12}$ & $0.304_{-0.012}^{+0.012}$ & $0.304_{-0.012}^{+0.013} $\\
$\sin ^{2} \theta_{23}$ & $0.573_{-0.020}^{+0.016}$ & $0.575_{-0.019}^{+0.016}$ \\
$\sin ^{2} \theta_{13}$ & $0.02219_{-0.00063}^{+0.00062}$ & $0.02238_{-0.00062}^{+0.00063}$ \\
$\frac{\Delta m_{21}^{2}}{10^{-5} \mathrm{\,eV}^{2}}$ & $7.42_{-0.20}^{+0.21}$ & $7.42_{-0.20}^{+0.21}$ \\
$\frac{\Delta m_{3l}^{2}}{10^{-3} \mathrm{\,eV}^{2}}$ & $+2.517_{-0.028}^{+0.026}$ & $-2.498_{-0.028}^{+0.028}$ \\
\hline
\end{tabular}
\centering
\caption{\label{table} The best fit parameters (BFP) and $1\sigma$ allowed ranges of the neutrino oscillation parameters for both the normal ordering (NO) and inverted ordering (IO) scenarios, derived from a global fit of the Super-Kamiokande atmospheric neutrino oscillation data by Nufit\,\cite{Esteban:2020cvm}. Note that $\Delta m_{3l}^2\equiv\Delta m_{31}^2>0$ for NO, and $\Delta m_{3l}^2\equiv \Delta m_{32}^2<0$ for IO.}
\end{table}

Currently the absolute value of neutrino masses have not been measured, and the strongest constraint is obtained from cosmological observation, $\Sigma m_i<0.12\,\text{eV}$\,\cite{Planck:2018vyg}. In order to generate the observed neutrino masses successfully, while also ensuring that the Yukawa couplings remain perturbative, it is required that $v_\Delta\gtrsim 0.05\text{\,eV}$\,\cite{Barrie:2022cub}. 

After EWSB, the massless gauge bosons acquire masses through the Higgs mechanism, and the 10 degrees of freedom in the scalar sector reduces to 7 physical scalar fields with definite mass: $H^{\pm\pm},H^{\pm},H^0,A^0,h^0$. The $H^{\pm\pm}$ is simply $\Delta^{\pm\pm}$, the $H^{\pm}$ is given by the mixing of $\Delta^{\pm}$ and $\phi^{\pm}$, and the mixing of $\Delta^0$ and $\phi^0$ gives rise to $H^0,A^0$ and standard model Higgs $h^0$\,\cite{Mandal:2022zmy,Arhrib:2011uy,Bonilla:2015eha}.

\subsection{Affleck-Dine leptogenesis through type II seesaw}

Since the minimal type II seesaw model can not lead to successful thermal leptogenesis, the Affleck-Dine mechanism is taken into account. In the Affleck-Dine mechanism\,\cite{Affleck:1984fy}, a scalar field carrying nonzero baryon or lepton number acquire a large vacuum expectation value along the flat direction of the potential in the early universe. With baryon or lepton number violating terms in the potential, the nontrivial angular motion in the phase of the scalar field during the evolution of universe leads to the generation of baryon or lepton number asymmetry. Fortunately, the scalar field carrying lepton number required in the Affleck-Dine mechanism can be provided by the triplet Higgs $\Delta$. Furthermore, the non-minimal coupling of the scalars $H$ and $\Delta$ to gravity induces a flat potential required by the Affleck-Dine mechanism and a Starobinsky type inflation during the early universe\,\cite{Starobinsky:1980te,Bezrukov:2007ep,Barbon:2009ya,Bezrukov:2010jz,Hertzberg:2010dc,Burgess:2010zq,Giudice:2010ka,Lebedev:2011aq}. This framework is called type II seesaw leptogenesis\,\cite{Barrie:2021mwi,Barrie:2022cub,Han:2022ssz}, and in this framework, the origin of neutrino mass, inflation and baryon asymmetry may be explained simultaneously.

Considering the non-minimal couplings of $\Delta$ and $H$ to gravity, the relevant Lagrangian in Jordan frame can be written as
\begin{equation}
    \frac{\mathcal{L}}{\sqrt{-g}}\supset -\frac{1}{2}M^2_P R - f(H,\Delta) R + g^{\mu\nu}(D_\mu H)^\dagger(D_\nu H) + g^{\mu\nu}\text{Tr}(D_\mu \Delta)^\dagger(D_\nu \Delta) - V(H,\Delta)
\end{equation}
where $R$ is Ricci scalar. To simplify the analysis, we focus on the neutral components $\phi^0$ and $\delta^0$ and consider the non-minimal coupling to be
\begin{equation}
    f(H,\Delta)=\xi_H |\phi^0|^2 + \xi_\Delta |\Delta^0|^2
\end{equation}
After a Weyl transformation, the Lagrangian can be written in Einstein frame, in which the gravitational portion is of Einstein-Hilbert form. The scalar potential in Einstein frame is
\begin{equation}
    V_E(H,\Delta)=\frac{M_P^4}{\left(M_P^2+2f(H,\Delta)\right)^2}V(H,\Delta)
\end{equation}
which exhibits a flat direction at the large field limit of $\phi^0$ and $\Delta^0$. This flat direction can be recognized as a Starobinsky-like inflationary trajectory, and the inflaton is the mixing of $\Delta^0$ and $\phi^0$. 

During the inflationary evolution, the nontrivial motion of the angular direction of $\Delta^0$ generate nonzero lepton number, which subsequently transfer to ordinary particles during reheating. After reheating and before electroweak phase transition, a part of net lepton number is converted to baryon number through the sphaleron process\,\cite{Kuzmin:1985mm,Harvey:1990qw,Klinkhamer:1984di,Trodden:1998ym}.

However, any lepton number violating processes in thermal equilibrium after reheating will wash out the generated lepton asymmetry. Thus it is required that the processes $L L \leftrightarrow H H$ and $H H \leftrightarrow \Delta$ are never in thermal equilibrium, 
\begin{equation}
\left.\Gamma\right|_{T=m_{\Delta}}=n\langle\sigma v\rangle \approx h^{2} \mu^{2} / m_{\Delta}<\left.H\right|_{T=m_{\Delta}}
\end{equation}
\begin{equation}
\label{eq:washout_condition}
    \left.\Gamma_{I D}(H H \leftrightarrow \Delta)\right|_{T=m_{\Delta}}
    \simeq \frac{\mu^{2}}{32 \pi m_{\Delta}}<\left.H\right|_{T=m_{\Delta}}
\end{equation}
where $\left.H\right|_{T=m_{\Delta}}=\sqrt{\frac{\pi^{2} g_{*}}{90}} \frac{m_{\Delta}^{2}}{M_{p}}$. Using $v_{\Delta} \simeq\frac{\mu v_{\mathrm{EW}}^{2}}{2 m_{\Delta}^{2}}$ and Eq.\eqref{eq:washout_condition}, the necessary condition to avoid washout effect is found to be
\begin{equation}
    v_{\Delta} \lesssim 10^{-5} \mathrm{\,GeV}\left(\frac{m_{\Delta}}{1 \mathrm{\,TeV}}\right)^{-1 / 2}.
\end{equation}

For $m_{\Delta} \gtrsim 1 \mathrm{\,TeV}$\,\cite{ATLAS:2017xqs}, we require that $v_{\Delta} \lesssim 10 \mathrm{\,keV}$ to prevent the washout effect and achieve successful leptogenesis. In the limit of $v_\Delta\ll v_{EW}$ and $m_\Delta\gg v_{EW}$, the mass spectrum of the scalar sector except for the SM Higgs is approximately degenerate,
\begin{equation}
    m_{H^{\pm\pm}}\simeq m_{H^{\pm}}\simeq m_{H^0/A^0}\simeq m_{\Delta}.
\end{equation}

\section{The Lepton Flavor Violation Processes}
\label{section3}
In the type II seesaw model the couplings of triplet Higgs scalar with the SM lepton sector lead to charged lepton flavor violation processes, and the most stringent constraints from experiments are set by the muon sector, including $\mu\to e\gamma$, $\mu\to3e$ and $\mu\to e$ conversion in nuclei\,\cite{Dinh:2012bp,Ardu:2022sbt,Frau:2024rzt}. The relevant parameters include the Yukawa couplings, the mass scale of triplet scalar $m_\Delta$ and the lepton number violating coupling $\mu$. Note that the Yukawa couplings can be parameterized by neutrino masses and the PMNS matrix, which are relevant in neutrino oscillation experiments. On the other hand, the other two parameters are relevant to leptogenesis. Thus, in this section, we conduct an integral analysis of both parts of parameters and figure out that the constraints from CLFV experiments with the requirement of satisfying the current neutrino oscillation data lead to constraints on the parameters relevant to leptogenesis. Since the neutrino oscillation experiments have not pinned down all the parameters of neutrino masses and mixing, we adopt the Monte Carlo method to scan the parameter region to satisfy the neutrino data within 3$\sigma$ range. Finally we obtain the most conservative constraints on the parameters of type II seesaw leptogenesis consistent with current neutrino data, and also obtain the future experimental reach of the parameter space.

In type II seesaw the leading contribution to $\mu\rightarrow 3e$ amplitude is at tree level, while the contribution to $\mu\rightarrow e\gamma$ and $\mu\rightarrow e$ conversion arise at one-loop. The corresponding effective low energy LFV Lagrangian generated at one-loop contributing to the $\mu$-$e$ transition processes can be written as\,\cite{Dinh:2012bp}:
\begin{eqnarray}
\label{effectiveLagrangian}
\mathcal{L}^{\text{eff}}&=&-4\frac{eG_F}{\sqrt{2}}\left(m_\mu A_R \bar{e}\sigma^{\alpha\beta}P_R\mu F_{\beta\alpha}+h.c.\right)\nonumber\\
&&-\frac{e^2G_F}{\sqrt{2}}\left(A_L\left(-m_\mu^2\right)\bar{e}\gamma^\alpha P_L\mu\sum_{Q=u,d}q_Q\bar{Q}\gamma_\alpha Q+h.c.\right)~,
\end{eqnarray}
where $e$ is the proton charge, and $q_Q$ is the electric charge of the up or down quark. The form factors $A_{R,L}$ are given by,
\begin{equation}
\label{AR}
    A_R=-\frac{1}{\sqrt{2}G_F}\frac{\left(h^\dagger h\right)_{e\mu}}{48\pi^2}\left(\frac{1}{8m^2_{H^+}}+\frac{1}{m^2_{H^{++}}}\right)~,
\end{equation}
\begin{equation}
\label{AL}
    A_L\left(q^2\right)=-\frac{1}{\sqrt{2}G_F}\frac{h_{le}^*h_{l\mu}}{6\pi^2}\left(\frac{1}{12m_{H^+}^2}+\frac{1}{m^2_{H^{++}}}f\left(\frac{-q^2}{m^2_{H^{++}}},\frac{m^2_l}{m^2_{H^{++}}}\right)\right)~,
\end{equation}
where $m_l$ is the mass of the charged lepton $l, l=e,\mu,\tau$. The loop function $f(r,s_l)$ is well known:
\begin{equation}
\label{loopfunction}
    f\left(r,s_l\right)=\frac{4s_l}{r}+\log(s_l)+\left(1-\frac{2s_l}{r}\right)\sqrt{1+\frac{4s_l}{r}}\log\frac{\sqrt{r}+\sqrt{r+4s_l}}{\sqrt{r}-\sqrt{r+4s_l}}~.
\end{equation}
In the limit in which the charged lepton masses $m_l\ll m_{H^{++}}$, one has $f(r,s_l)\simeq\log r=\log(m_\mu^2/m^2_{H^{++}})$.

\subsection{The $\mu\to e\gamma$ Decay}

Using Eq.\eqref{effectiveLagrangian} and Eq.\eqref{AR} we get the branching ratio of $\mu\to e\gamma$ decay:
\begin{equation}
\label{branchingratio1}
    \text{BR}\left(\mu\to e\gamma\right)\cong 384\pi^2(4\pi\alpha_{em})\left|A_R\right|^2=\frac{\alpha_{em}}{192\pi}\frac{\left|\left(h^\dagger h\right)_{e\mu}\right|^2}{G^2_F}\left(\frac{1}{m^2_{H^+}}+\frac{8}{m^2_{H^{++}}}\right)^2~,  
\end{equation}
where $\alpha_{em}$ is the fine structure constant, and $G_F$ is the Fermi constant.

\begin{figure}
    \centering
    \includegraphics[width=0.8\textwidth]{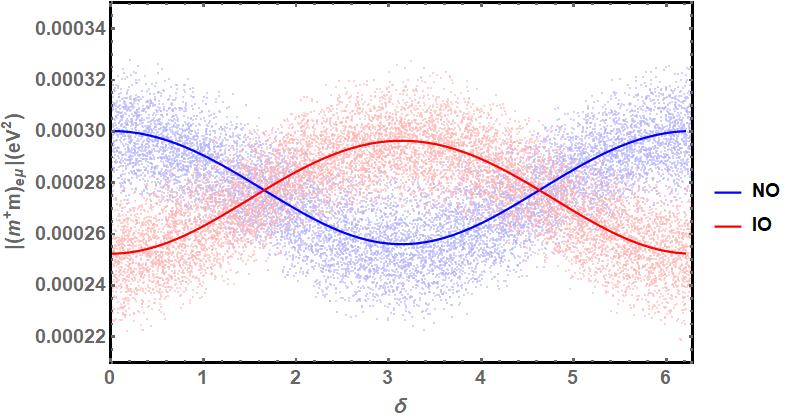}
    \caption{The $\left|\left(m^\dagger m\right)_{e\mu}\right|$ parameter, being independent of the lightest neutrino mass and Majorana phases, as a function of Dirac phase for the NO (Blue) and IO (Red) scenarios once the mixing angles and mass differences are fixed by the best fit values in Table\,\ref{table}. The scattered points are obtained by varying the neutrino oscillation parameters within their corresponding $3\sigma$ intervals and giving random values to the Dirac phase.}
    \label{fig:mutoegamma}
\end{figure}

At present, the strongest constraint on this process is given by the MEG collaboration\,\cite{Adam:2013vqa}, with an upper limit on the branching ratio,
\begin{equation}
\label{MEG}
    \text{BR}\left(\mu\to e\gamma\right)<4.2\times 10^{-13},\quad 90\%\text{C.L}.
\end{equation}
For $m_{H^+}\cong m_{H^{++}}\equiv m_\Delta$, using Eq.\eqref{branchingratio1} and Eq.\eqref{MEG} we can derive a bound on the Yukawa couplings,
\begin{equation}
    \left|\left(h^\dagger h\right)_{e\mu}\right|<1.6\times10^{-4}\left(\frac{m_\Delta}{800\text{\,GeV}}\right)^2~,
\end{equation}
from which we can derive the following $m_\Delta$ dependent lower bound on the triplet Higgs parameter $\mu$,
\begin{equation}
\label{eq_meg_mu}
\mu>1.7\times10^{-6}\text{\,GeV}\frac{\sqrt{\left|\left(m^\dagger m\right)_{e\mu}\right|}}{1\text{\,eV}}\frac{m_\Delta}{800\text{\,GeV}}~,
\end{equation}
where we have used Eq.\eqref{tripletHiggsvev}. From Eq.\eqref{Yukawacouplings} it is not difficult to get:
\begin{eqnarray}
\label{mdaggermemu}
    \left|\left(m^\dagger m\right)_{e\mu}\right|&=&\left|c_{13}\left(\Delta m_{21}^2c_{12}c_{23}s_{12}+e^{-i\delta}\left(\Delta m_{31}^2-\Delta m_{21}^2s_{12}^2\right)s_{13}s_{23}\right)\nonumber\right|\\
    &=&\left|U_{e2}U^\dagger_{2\mu}\Delta m^2_{21}+U_{e3}U^\dagger_{3\mu}\Delta m^2_{31}\right|~,
\end{eqnarray}
where $c_{ij}=\cos\theta_{ij}$ and $s_{ij}=\sin\theta_{ij}$. From the above equation we can see that $\left|\left(m^\dagger m\right)_{e\mu}\right|$ is independent of the Majorana neutrino phases and the unknown absolute neutrino masses, varying only with the $\delta$ phase once the mixing angles and mass differences are fixed (see Fig.\,\ref{fig:mutoegamma}). 

As illustrated by the scattered points in Fig.\,\ref{fig:mutoegamma}, we scanned the whole neutrino mass and mixing parameters consistent with neutrino oscillation experiments with Monte Carlo method. The scattered points are obtained by varying all the neutrino oscillation parameters within the corresponding $3\sigma$ intervals and allowing for arbitrary values of the Dirac phase in $\left[0,2\pi\right]$ with a uniform distribution of random sampling. In order to obtain the most conservative constraints, the minimum of $\left|\left(m^\dagger m\right)_{e\mu}\right|$ obtained in Monte Carlo simulation are adopted,
\begin{equation}
    \left|\left(m^\dagger m\right)_{e\mu}\right|_{\text{min}}\sim 2.2\times 10^{-4}\text{\,eV}^2~,
\end{equation}
which is almost the same for NO and IO. Substitute the above result into Eq.\eqref{eq_meg_mu}, we get
\begin{equation}  \mu>2.5\times10^{-8}\text{\,GeV}\frac{m_\Delta}{800\text{\,GeV}}~,
\end{equation}
as the Green lines shown in the Fig.\,\ref{fig:result_no} and Fig.\,\ref{fig:result_io}.

The upcoming experiment MEG II, which is an upgrade of the MEG experiment, promises to reach a sensitivity of\,\cite{MEGII:2018kmf}
\begin{equation}
\label{MEG}
    \text{BR}\left(\mu\to e\gamma\right)<6\times 10^{-14},.
\end{equation}
which is one order of magnitude better than the current limit. The expected sensitivity on the parameter space is
\begin{equation}  \mu>4.9\times10^{-8}\text{\,GeV}\frac{m_\Delta}{800\text{\,GeV}}~,
\end{equation}
as the Green dashed lines shown in the Fig.\,\ref{fig:result_no} and Fig.\,\ref{fig:result_io}.

\subsection{The $\mu\to 3e$ Decay}

\begin{figure}
    \centering
    \includegraphics[width=0.8\textwidth]{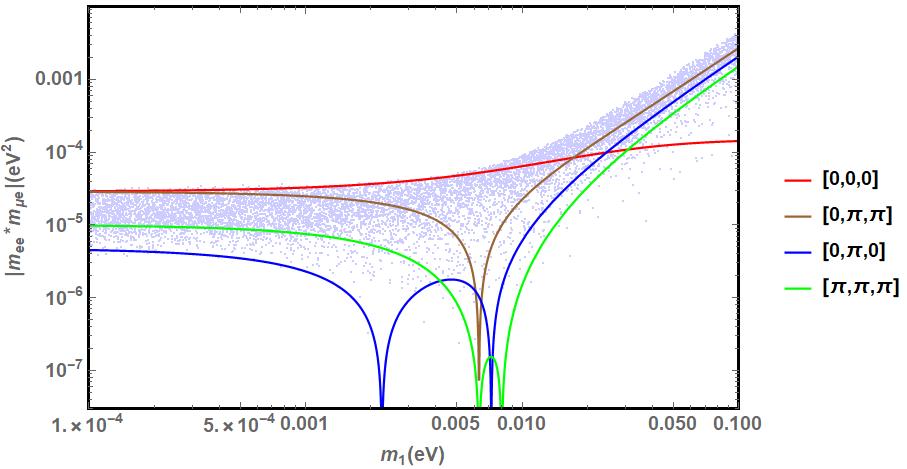}
    \caption{The dependence of $\left|m^*_{ee}m_{\mu e}\right|$ on the lightest neutrino mass $m_1$ in the NO scenario, for four sets of values of the Dirac and the two Majorana phases, $[\delta,\alpha_{21},\alpha_{31}]$. All the best fit values of the neutrino oscillation parameters are given in Table\,\ref{table}. The scattered points are obtained by varying the neutrino oscillation parameters within their corresponding $3\sigma$ intervals and giving random values to the Dirac and Majorana phases.}
    \label{fig:muto3e-no}
\end{figure}

\begin{figure}
    \centering
    \includegraphics[width=0.8\textwidth]{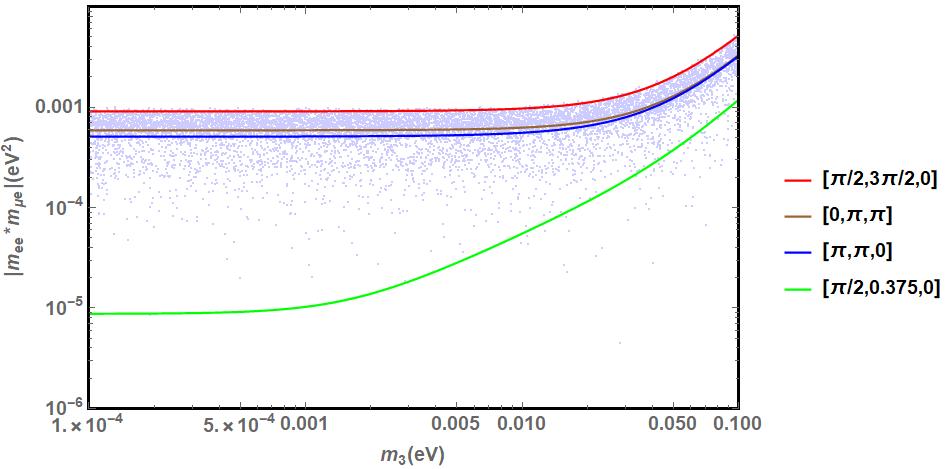}
    \caption{The same as in Fig.\,\ref{fig:muto3e-no} in IO scenario.}
    \label{fig:muto3e-io}
\end{figure}

The leading order contribution to the $\mu\to 3e$ decay amplitude is at the tree level, generated by the diagram with exchange of a virtual doubly-charged Higgs boson $\Delta^{++}$. The branching ratio can be easily calculated:
\begin{equation}
\label{branchingratio2}
    \text{BR}\left(\mu\to 3e\right)=\frac{1}{G^2_F}\frac{\left|\left(h^\dagger\right)_{ee}\left(h\right)_{\mu e}\right|^2}{m^4_{\Delta^{++}}}=\frac{1}{G_F^2m^4_{\Delta^{++}}}\frac{\left|m^*_{ee}m_{\mu e}\right|^2}{16v^4_\Delta}~,
\end{equation}
where the last step we have used Eq.\eqref{neutrinomass}, and where
\begin{eqnarray}
\label{eq:muto3e}
    \left|m^*_{ee}m_{\mu e}\right|&=&c_{13}\left[m_1c_{12}^2c_{13}^2+e^{i\alpha_{21}}m_2c_{13}^2s_{12}^2+e^{i(\alpha_{31}-2\delta)}m_3s_{13}^2\right][e^{-i(\alpha_{31}-\delta)}m_3s_{13}s_{23}\nonumber\\
    &&-m_1c_{12}(c_{23}s_{12}+e^{-i\delta}c_{12}s_{13}s_{23})+e^{-i\alpha_{21}}m_2s_{12}(c_{12}c_{23}-e^{-i\delta}s_{12}s_{13}s_{23})]~.
\end{eqnarray}
From the above equation we can also see that $\left|m^*_{ee}m_{\mu e}\right|$ is dependent of the lightest neutrino mass, Dirac phase and Majorana neutrino phases once the mixing angles and mass differences are fixed (see Fig.\,\ref{fig:muto3e-no} and Fig.\,\ref{fig:muto3e-io}).

Firstly, we consider the dependence of  $\left|m_{ee}\right|$ and CP phases. To do this we consider the CP phase parameter sets that maximize and minimize the $\left|m^*_{ee}m_{\mu e}\right|$ parameter. In the NO scenario, taking $m_1\lesssim10^{-4}$\,eV, the $\left|m_{ee}\right|$ component is found to be minimized when the CP phases satisfy $\alpha_{21}-\alpha_{31}+2\delta=\pi$, and maximized when this relation is equal to $0$. However, for $m_1\gtrsim10^{-4}$\,eV, one can have $\left|m_{ee}\right|=0$ for specific values of $m_1$ if CP phases $\alpha_{21}$ and $\alpha_{31}-2\delta$ satisfy that $\alpha_{21}=\pi$, and $\left(\alpha_{31}-2\delta\right)=0$ or $\pi$. This is illustrated in Fig.\,\ref{fig:muto3e-no}, for $\left[\delta_{CP},\alpha_{21},\alpha_{31}\right]=\left[0,\pi,0\right]$ we have $\left|m_{ee}\right|=0$ at $m_1\simeq2.25\times10^{-3}$\,eV, while in the case of the $\left[0,\pi,\pi\right]$ we have $\left|m_{ee}\right|=0$ at $m_1\simeq6.31\times10^{-3}$\,eV. 

If the light neutrino mass spectrum is with inverted ordering $\left(m_3\ll m_1<m_2\right)$, we have $\left|m_{ee}\right|\gtrsim\sqrt{\left|\Delta m_{31}^2\right|+m_3^2}\cos{2\theta_{12}}\gtrsim1.96\times10^{-2}$\,eV with $m_3=0$ for the best fit parameters. In addition, the appearance of zeros is also caused by the variation of the $\left|m_{\mu e}\right|$ component with the CP phases. In the NO scenario, the minimal value of $\left|m_{\mu e}\right|$ is obtained for $\alpha_{21}=\pi$, $\delta=0$ or $\pi$ and $\alpha_{31}=0$ or $\pi$. This can be shown in Fig.\,\ref{fig:muto3e-no}, for $\left[0,\pi,0\right]$, the zero takes place at $m_1\simeq7.2\times10^{-3}$\,eV, while for $\left[\pi,\pi,\pi\right]$ we have $\left|m_{\mu e}\right|=0$ at $m_1\simeq8.0\times10^{-3}$\,eV. On the other hand, the maximal value of $\left|m_{\mu e}\right|$ satisfies that $\alpha_{31}-\alpha_{21}=\delta$ and $\delta=\pi$, and we get $\left|m_{\mu e}\right|<8.1\times10^{-3}$\,eV. 

In the IO scenario with negligible $m_3\simeq0$, the maximal value of $\left|m_{\mu e}\right|$ corresponds to $\delta=0$ and $\alpha_{21}=\pi$ and is given by max $\left(\left|m_{\mu e}\right|\right)\simeq\sqrt{\left|\Delta m_{31}^2\right|}c_{13}(c_{23}\sin{2\theta_{12}}+s_{23}s_{13}\cos{2\theta_{12}})\simeq3.19\times10^{-2}$\,eV for the best fit parameters. The component $\left|m_{\mu e}\right|$ is strongly suppressed (see Fig.\,\ref{fig:muto3e-io}), i.e., we have $\left|m_{\mu e}\right|\ll\text{max}(\left|m_{\mu e}\right|)$, when $\delta\simeq\pi/2$ and the Majorana phase $\alpha_{21}$ which is determined by the equation,
\begin{equation}
\label{suppression-muto3e}
c_{23}c_{12}s_{12}\sin{\alpha_{21}}\simeq\left(c_{12}^2+s_{12}^2\cos{\alpha_{21}}\right)s_{23s_{13}}~,
\end{equation}
from which we derive the value $\alpha_{21}\simeq0.375$ for the best fit parameters in Table \ref{table}.

The properties of the $|m_{ee}|$ and $|m_{\mu e}|$ described above allow us to understand most of the specific features of the dependence of $\left|m^*_{ee}m_{\mu e}\right|$ on the neutrino mass ordering and the CP phases. In the NO scenario with negligible $m_1\simeq0$, the maximum of the $\left|m^*_{ee}m_{\mu e}\right|$ is obtained for $\alpha_{31}-\alpha_{21}=\delta=0$, that is,  $\left[\delta,\alpha_{21},\alpha_{31}\right]=\left[0,0,0\right]$, and Eq.\eqref{eq:muto3e} becomes
\begin{equation}
\label{max_meemmue_no}
    \text{max}\left(\left|m^*_{ee}m_{\mu e}\right|\right)=\left|\left(m_2s_{12}^2c_{13}^2+m_3s_{13}^2\right)c_{13}\left(m_2s_{12}\left(c_{12}c_{23}-s_{12}s_{23}s_{13}\right)+m_3s_{23}s_{13}\right)\right|~,
\end{equation}
where the best fit parameters of the neutrino oscillation parameters is given in Table \ref{table}, from which we can get the maximum of $\left|m^*_{ee}m_{\mu e}\right|$ is approximate to $2.81\times10^{-5}\text{\,eV}^2$.

On the other hand, in the IO scenario with $m_3\simeq0$, the maximum of the $\left|m^*_{ee}m_{\mu e}\right|$ is achieved when $\delta=0$ and $\alpha_{21}=\pi$, and reads:
\begin{equation}
\label{max_meemmue_io}
    \text{max}(\left|m^*_{ee}m_{\mu e}\right|)\simeq\left|\Delta m_{31}^2\right|c_{13}^2\left(\frac{1}{2}c_{23}\sin{4\theta_{12}}+s_{23}s_{13}\cos^2{2\theta_{12}}\right)\simeq6.1\times10^{-4}\text{\,eV}^2~,
\end{equation}
while the minimal case occurs at the parameter set satisfying Eq.\eqref{suppression-muto3e}, from which we can get the minimal value of the $\left|m^*_{ee}m_{\mu e}\right|\simeq9.9\times10^{-6}\text{\,eV}^2$. 

Finally, the scattered points in Fig.\,\ref{fig:muto3e-no} and Fig.\,\ref{fig:muto3e-io} are obtained by varying all the neutrino oscillation parameters within the corresponding $3\sigma$ intervals and allowing for arbitrary values of the Dirac and Majorana phases in $\left[0,2\pi\right]$. The most conservative constraints from $\mu\rightarrow3e$ process to our model can be derived from the minimal value of $\left|m_{ee}^*m_{\mu e}\right|$. In the IO scenario we obtain 
\begin{equation}
\left|m_{ee}^*m_{\mu e}\right|_{\text{min}}(\text{IO})\simeq1.1\times 10^{-5}\text{\,eV}^2.    
\end{equation} 
In the NO scenario, however, in some parameter region $\left|m_{ee}^*m_{\mu e}\right|$ may vanish, thus this part of parameter region is not constrained by $\mu\rightarrow3e$ experiments. Therefore in the NO scenario we focus on the parameter region with $m_1\lesssim10^{-3}\,$eV, and obtain
\begin{equation}
    \left|m_{ee}^*m_{\mu e}\right|_{\text{min}}(\text{NO})\simeq 2.4\times 10^{-6}\text{\,eV}^2,\quad m_1\lesssim10^{-3}\text{\,eV}.
\end{equation}


The current best upper bound on the $\mu\to 3e$ branching ratio is derived from the results of the SINDRUM experiment,\,\cite{SINDRUM:1987nra}
\begin{equation}
\label{SINDRUM}
    \text{BR}\left(\mu\to3e\right)<10^{-12},\quad90\%\text{C.L.},
\end{equation}
from the present limit and Eq.\eqref{branchingratio2}, one can obtain the following constraint on $\left|\left(h^\dagger\right)_{ee}\left(h\right)_{\mu e}\right|$:
\begin{equation}
\left|\left(h^\dagger\right)_{ee}\left(h\right)_{\mu e}\right|<7.5\times10^{-6}\left(\frac{m_\Delta}{800\text{\,GeV}}\right)^2~,
\end{equation}
from which and using Eq.\eqref{tripletHiggsvev} we also get the lower bound on the cubic coupling $\mu$,
\begin{equation}
\label{sindrum-mu}
\mu>7.7\times10^{-6}\text{\,GeV}\frac{\sqrt{\left|m^*_{ee}m_{\mu e}\right|}}{1\text{\,eV}}\frac{m_\Delta}{800\text{\,GeV}}~.
\end{equation}
Using Eq.\eqref{sindrum-mu} and the results of Monte Carlo, we determine the lower bounds on coupling term $\mu$ from the current SINDRUM experiment,
\begin{equation}
\label{eq_sindrum_mu}
\mu>1.2\left(2.6\right)\times10^{-8}\text{\,GeV}\frac{m_\Delta}{800\text{\,GeV}}~,
\end{equation}
where the bracket in the above equation means the case of IO. Note that from the above equation, we can get the parameter space between the cubic coupling $\mu$ term and the mass of triplet Higgs from the $\mu\to3e$ decay process in the case of the normal ordering and inverted ordering for the neutrino mass, as we shown the Red lines in Fig.\,\ref{fig:result_no} and Fig.\,\ref{fig:result_io}.

The upcoming experiment known as Mu3e promises to deliver significantly improved sensitivity to the $\mu\to3e$ process, potentially probing the branching ratio by an additional four orders of magnitude, that is\,\cite{Mu3e:2020gyw}
\begin{equation}
    \text{BR}_{\text{Mu3e}}\left(\mu\to3e\right)<10^{-16}~,
\end{equation}
which will provide the corresponding limit on the Yukawa couplings,
\begin{equation}
\left|\left(h^\dagger\right)_{ee}\left(h\right)_{\mu e}\right|<7.5\times10^{-8}\left(\frac{m_\Delta}{800\text{\,GeV}}\right)^2~,
\end{equation}
and the cubic coupling $\mu$,
\begin{equation}
\label{mu3e-mu}
\mu>7.7\times10^{-5}\text{\,GeV}\frac{\sqrt{\left|m^*_{ee}m_{\mu e}\right|}}{1\text{\,eV}}\frac{m_\Delta}{800\text{\,GeV}}~,
\end{equation}
providing an order of magnitude improvement to the sensitivity to $\mu$.

Similarly, using Eq.\eqref{mu3e-mu} and Monte Carlo results, the future expected lower bounds on the cubic coupling $\mu$ from Mu3e experiment can be determined,
\begin{equation}
\label{eq_mu3e_mu}
\mu>1.2\left(2.6\right)\times10^{-7}\text{\,GeV}\frac{m_\Delta}{800\text{\,GeV}}~,
\end{equation}
where the bracket means the case of IO.
Similarly, from Eq.\eqref{eq_mu3e_mu}, we can also get the parameter space between the cubic coupling $\mu$ term and the mass of triplet Higgs from the $\mu\to3e$ decay process in the case of the normal ordering and inverted ordering for the neutrino mass, as we shown the Red dashed lines in Fig.\,\ref{fig:result_no} and Fig.\,\ref{fig:result_io}.

\subsection{The $\mu-e$ Conversion in Nuclei}

Finally we consider the $\mu-e$ conversion in a generic nucleus $\mathcal{N}$. From the interaction Lagrangian Eq.(\ref{effectiveLagrangian}), we get the conversion rate of the $\mu\to e$ conversion in Nuclei process in the type-II seesaw scenario
\begin{equation}
\label{conversionrate}
    \text{CR}\left(\mu\mathcal{N}\to e\mathcal{N}\right)\cong\left(4\pi\alpha_{em}\right)^2\frac{2G_F^2}{\Gamma_{\text{capt}}}\left|A_R\frac{D}{\sqrt{4\pi\alpha_{em}}}+\left(2q_u+q_d\right)A_LV^{(p)}\right|^2~,
\end{equation}
where $A_R$ and $A_L$ have been presented in Eq.\eqref{AR} and Eq.\eqref{AL}. In Eq.\eqref{conversionrate}, the parameters $D$ and $V^{(p)}$ present overlap integrals of the muon and electron wave functions, and $\Gamma_\text{capt}$ is the experimentally known total muon capture rate\,\cite{Kitano:2002mt}. 

Considering a light nucleus, one has with a good approximation $D\simeq8\sqrt{4\pi\alpha_{em}}V^{(p)}$, with the vector type overlap integral of the proton, $V^{(p)}$, given by
\begin{equation}
\label{Vp}
    V^{(p)}\simeq\frac{1}{4\pi}m^{5/2}_\mu\alpha_{em}^{3/2}Z^2_{eff}Z^{1/2}F\left(-m_\mu^2\right)~,
\end{equation}
where $F(q^2)$ is the nuclear form factor at momentum transfer squared $q^2=-m_\mu^2$, $m_\mu$ being the muon mass, and where $Z_{eff}$ is an effective atomic charge. Using the result for $D$ quoted above, Eq.\eqref{AR}, Eq.\eqref{AL} and Eq.\eqref{Vp}, the conversion rate can be written as
\begin{equation}
\begin{aligned}
    \text{CR}\left(\mu\mathcal{N}\to e\mathcal{N}\right)&\cong\frac{\alpha_{em}^5}{36\pi^4}\frac{m_\mu^5}{\Gamma_{\text{capt}}}Z_{eff}^4ZF^2(-m_\mu^2)\times\\
    &\left|\left(h^\dagger h\right)_{e\mu}\left[\frac{5}{24m^2_{H^+}}+\frac{1}{m^2_{H^{++}}}\right]+\frac{1}{m^2_{H^{++}}}\sum_{l=e,\mu,\tau}h^\dagger_{el}f\left(r,s_l\right)h_{l\mu}\right|^2~.
\end{aligned}
\end{equation}

The two nuclei that were used in the past and are of interest in the future in $\mu-e$ conversion experiments are ${}^{48}_{22}\text{Ti}$ and ${}^{27}_{13}\text{Al}$. The relevant parameters involved in the calculation of $\mu~-~e$ conversion are given in Table\,\ref{table2}.

\begin{table}
\renewcommand\arraystretch{1.2}
\tabcolsep=0.5cm
\begin{tabular}{|c|ccccc|}
\hline\hline $\mathcal{N}$ & $Z_{eff}$ & $D m_\mu^{-5/2}$ & $V^{(p)}m_\mu^{-5/2}$ & $\Gamma_\text{capt}(10^6s^{-1})$ & $F(-m_\mu^2)$\\
\hline ${}^{48}_{22}\text{Ti}$ & $17.6$ & $0.0864$ & $0.0396$ & $2.590$ & $0.54$\\
${}^{27}_{13}\text{Al}$ &$11.62$ & $0.0362$ & $0.0161$ & $0.7054$ & $0.64$\\
\hline\hline
\end{tabular}
\centering
\caption{\label{table2}Nuclear parameters relevant to $\mu-e$ conversion in ${}^{48}_{22}\text{Ti}$ and ${}^{27}_{13}\text{Al}$\,\cite{Kitano:2002mt,Dinh:2012bp}.}
\end{table}

\begin{figure}
    \centering
    \includegraphics[width=0.8\textwidth]{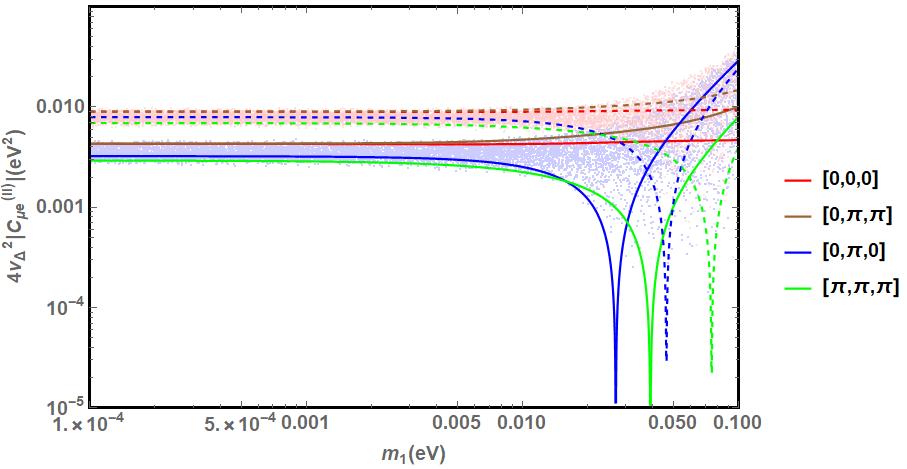}
    \caption{The dependence of $4v_\Delta^2\left|C_{\mu e}^{(II)}\right|$ on the lighest neutrino mass $m_1$ in the case of NO, for four sets of values of the Dirac and the two Majorana phases, $[\delta_{CP},\alpha_{21},\alpha_{31}]$ and $m_\Delta=800 (2\times 10^6)$\,GeV, plain (dashed) curves. The figure is obtained for the best fit values of the neutrino oscillation parameters given in Table\,\ref{table}. The scattered points are obtained by varying the neutrino oscillation parameters within their corresponding $3\sigma$ intervals and giving random values to the Dirac and Majorana phases.}
    \label{fig:mutoe-no}
\end{figure}

\begin{figure}
    \centering
    \includegraphics[width=0.8\textwidth]{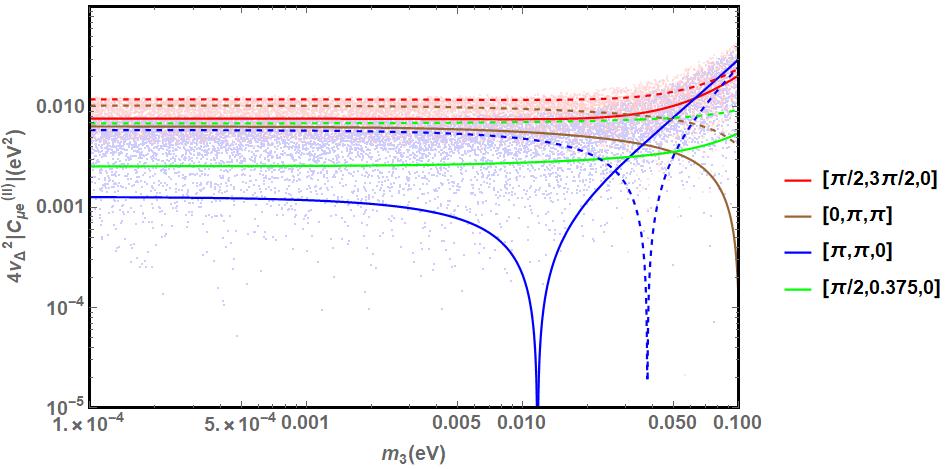}
    \caption{The same as in Fig.\,\ref{fig:mutoe-no} in the IO scenario.}
    \label{fig:mutoe-io}
\end{figure}

Thus, assuming $m_{H^+}\simeq m_{H^{++}}=m_\Delta$, we have $\text{CR}\left(\mu\mathcal{N}\to e\mathcal{N}\right)\propto\left|C_{\mu e}^{(II)}\right|^2$, where
\begin{equation}
\label{conversionpart}
    C_{\mu e}^{(II)}\equiv\frac{1}{4v_\Delta^2}\left[\frac{29}{24}\left(m^\dagger m\right)_{e\mu}+\sum_{l=e,\mu,\tau}m_{el}^\dagger f(r,s_l)m_{l\mu}\right]~,
\end{equation}
and we have used Eq.\eqref{neutrinomass}. In the above equation, the first term is given by Eq.\eqref{mdaggermemu}, and the form of $f(r,s_l)$ is depicted by Eq.\eqref{loopfunction}. Using Eq.\eqref{conversionpart}, we can get the dependence of  $4v_\Delta^2\left|C_{\mu e}^{(II)}\right|$ and the lightest neutrino mass, Dirac phase and Majorana phases in the PMNS matrix once the mixing angles and mass differences are fixed, as depicted in Fig.\,\ref{fig:mutoe-no} and Fig.\,\ref{fig:mutoe-io}.

For $m_\Delta=800\left(2\times10^6\right)$\,GeV and NO scenario with $m_1\ll10^{-3}$, the maximal value of $4v_\Delta^2\left|C_{\mu e}^{(II)}\right|$ takes place in $\left[\delta,\alpha_{21},\alpha_{31}\right]=\left[0,0,0\right]$ and at the maximum we have $4v_\Delta^2\left|C_{\mu e}^{(II)}\right|\simeq3.78\left(8.04\right)\times10^{-3}$\,eV (see Fig.\,\ref{fig:mutoe-no}) for the best fit parameters given in Table \ref{table}. 

As analyzed at Section \ref{section2}, the fact that the loop function is negative when $m_\Delta$ varying from $800$\,GeV to $2\times10^6$\,GeV allows us to get the suppression of the conversion rate. For values of the CP phases $\left[\delta,\alpha_{21},\alpha_{31}\right]=\left[0,\pi,0\right]$ and $m_\Delta=800\left(2\times10^6\right)$\,GeV,  $4v_\Delta^2\left|C_{\mu e}^{(II)}\right|$ goes through zero at $m_1\simeq0.0252\left(0.0439\right)$\,eV (see Fig.\,\ref{fig:mutoe-no}).

On the other hand, in the IO case with negligible $m_3$ ($m_3\ll10^{-3}$), the maximal value of $4v_\Delta^2\left|C_{\mu e}^{(II)}\right|$ occurs for $\left[\delta,\alpha_{21},\alpha_{31}\right]=\left[\pi/2,3\pi/2,0\right]$ and at the maximum in this case we have $4v_\Delta^2\left|C_{\mu e}^{(II)}\right|\simeq7.9\left(12.2\right)\times10^{-3}$ for $m_\Delta=800\left(2\times10^6\right)$\,GeV. While the minimum of $4v_\Delta^2\left|C_{\mu e}^{(II)}\right|$ for small $m_3$ can be found when the CP phases take $\left[\pi,\pi,0\right]$. At the minimum for $m_\Delta=800\left(2\times10^6\right)$\,GeV we get $4v_\Delta^2\left|C_{\mu e}^{(II)}\right|\simeq1.3\left(6.04\right)\times10^{-3}$\,eV. Taking the CP phase parameter set $\left[\pi,\pi,0\right]$, one can get a strong suppression of the conversion ratio for $m_3\simeq0.012\left(0.038\right)$\,eV and $m_\Delta=800\left(2\times10^6\right)$\,GeV (see Fig.\,\ref{fig:mutoe-io}). 

Finally, the scattered points in Fig.\,\ref{fig:mutoe-no} and Fig.\,\ref{fig:mutoe-io} are obtained by varying all the neutrino oscillation parameters within the corresponding $3\sigma$ intervals and allowing for arbitrary values of the Dirac and Majorana phases in $\left[0,2\pi\right]$. Since the value of $4v_\Delta^2\left|C_{\mu e}^{(II)}\right|$ vanishes at some part of parameter region, we consider only the region with $m_{1}(m_3)\lesssim10^{-3}\text{\,eV}$, and obtain $4v_\Delta^2\left|C_{\mu e}^{(II)}\right|_{\text{min}}\simeq 2.5\times10^{-3}(5.5\times10^{-5})\text{\,eV}^2$ for $m_\Delta=800\text{\,GeV}$ and  $4v_\Delta^2\left|C_{\mu e}^{(II)}\right|_{\text{min}}\simeq 6(3.9)\times10^{-3}\text{\,eV}^2$ for $m_\Delta=2\times10^6\text{\,GeV}$ in the NO(IO) scenario.

\subsubsection{The $\mu-e$ Conversion in Ti nuclei}

Currently, the best constraints on the conversion rate are provided by the SINDRUM II experiment, which utilised Ti nuclei,\,\cite{SINDRUMII:1993gxf}
\begin{equation}
    \text{CR}\left(\mu\mathcal{N}\to e\mathcal{N}\right)<4.3\times10^{-12},\quad 90\%\text{C.L.}.
\end{equation}
The constraint translates to the following limit,
\begin{equation}
    \left|C_{\mu e}^{(II)}\right|<7.94\times10^{-3}\left(\frac{m_\Delta}{800\text{\,GeV}}\right)^2~,
\end{equation}
with a corresponding limit on the cubic coupling $\mu$,
\begin{equation}
\label{eq_sindrum_ti_mu}
\mu>2.4\times10^{-7}\text{\,GeV}\frac{\sqrt{4v_\Delta^2\left|C_{\mu e}^{(II}\right|}}{1\text{\,eV}}\frac{m_\Delta}{800\text{\,GeV}}~,
\end{equation}
Using Eq.\eqref{eq_sindrum_ti_mu} and Monte Carlo method, at present the most conservative limits of $\mu\to e$ conversion in Ti nuclei can be obtained in the Fig.\,\ref{fig:result_no} and \ref{fig:result_io} as the Blue lines shown.

\subsubsection{The $\mu-e$ Conversion in Al Nuclei}
The upcoming COMET and Mu2e experiments, which search for $\mu-e$ conversion in the field of Al nuclei, is aiming to improve the current limit by several orders of magnitude,\,\cite{COMET:2018auw,Mu2e:2014fns} 
\begin{equation}
\begin{aligned}
    \text{CR}_\text{COMET}\left(\mu\mathcal{N}\to e\mathcal{N}\right)&<7\times10^{-15}~,\\
    \text{CR}_\text{Mu2e}\left(\mu\mathcal{N}\to e\mathcal{N}\right)&<6.2\times10^{-16}~.
\end{aligned}
\end{equation}

we consider the better sensitivity provided by Mu2e, which can be translated into the following sensitivity,
\begin{equation}
    \left|C_{\mu e}^{(II)}\right|<1.2\times10^{-4}\left(\frac{m_\Delta}{800\text{\,GeV}}\right)^2~,
\end{equation}
with a correspondingly a limit on the cubic coupling $\mu$,
\begin{equation}
\label{eq_mu2e_mu}
\mu>1.9\times10^{-6}\text{\,GeV}\frac{\sqrt{4v_\Delta^2\left|C_{\mu e}^{(II)}\right|}}{1\text{\,eV}}\frac{m_\Delta}{800\text{\,GeV}}.
\end{equation}
Therefore, as previously calculated, by the Monte Carlo method, the future sensitivities of $\mu\to e$ conversion in Al nuclei can be found in the Fig.\,\ref{fig:result_no} and Fig.\,\ref{fig:result_io} as the Purple dashed lines shown.

\subsection{Current constraints and future sensitivities}
\begin{figure}
    \centering
    \subfigure[Current Limits]{\includegraphics[height=0.23\textheight,width=0.49\textwidth]{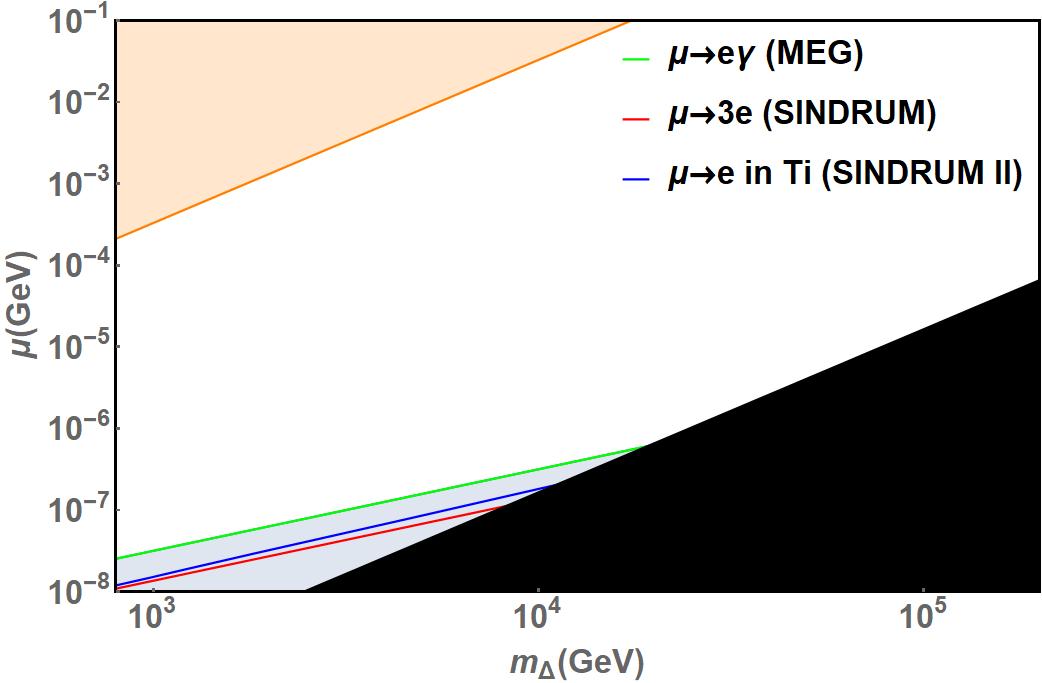}}
    \subfigure[Future Sensitivities]{\includegraphics[height=0.23\textheight,width=0.49\textwidth]{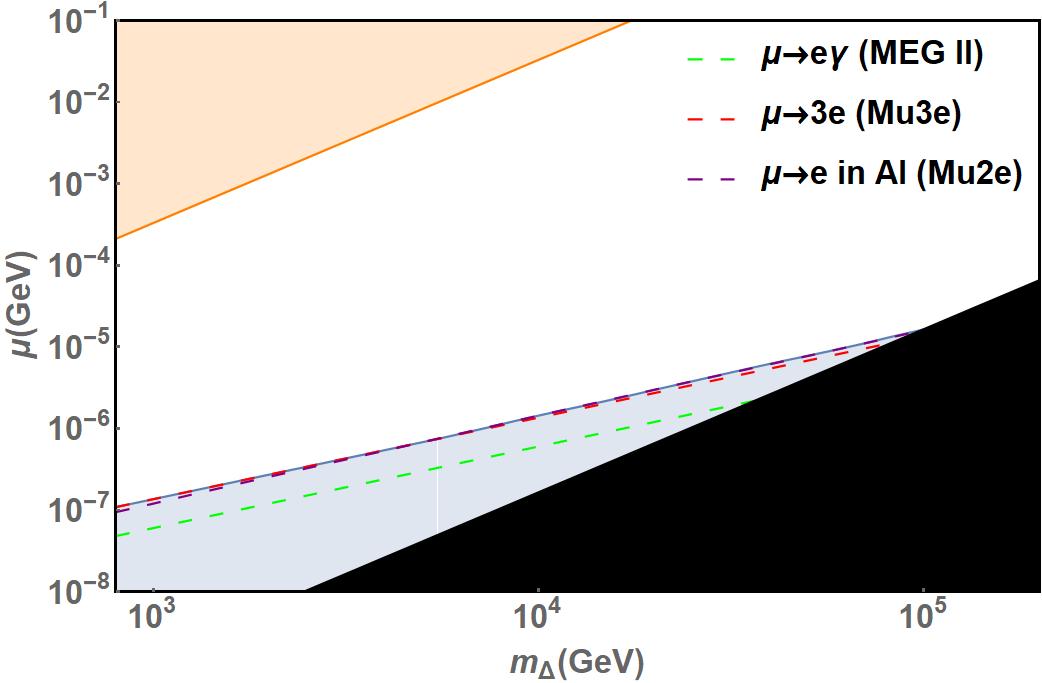}}
    \caption{\label{fig:result_no}The current limits (Left) and future experimental sensitivities (Right) is depicted for the most conservative values of $\left|\left(m^\dagger m\right)_{e\mu}\right|$, $\left|m_{ee}^\dagger m_{\mu e}\right|$ and $4v_\Delta^2\left|C_{\mu e}^{(II)}\right|$ in the NO scenario. The neutrino parameters are given by the best fit parameters in Table\,\ref{table} within their $3\sigma$ intervals. The constraints from requiring perturbative Yukawa couplings and avoiding the wash out effect are depicted as the black and orange region respectively.}
\end{figure}

\begin{figure}
    \centering
    \subfigure[Current Limits]{\includegraphics[height=0.23\textheight,width=0.49\textwidth]{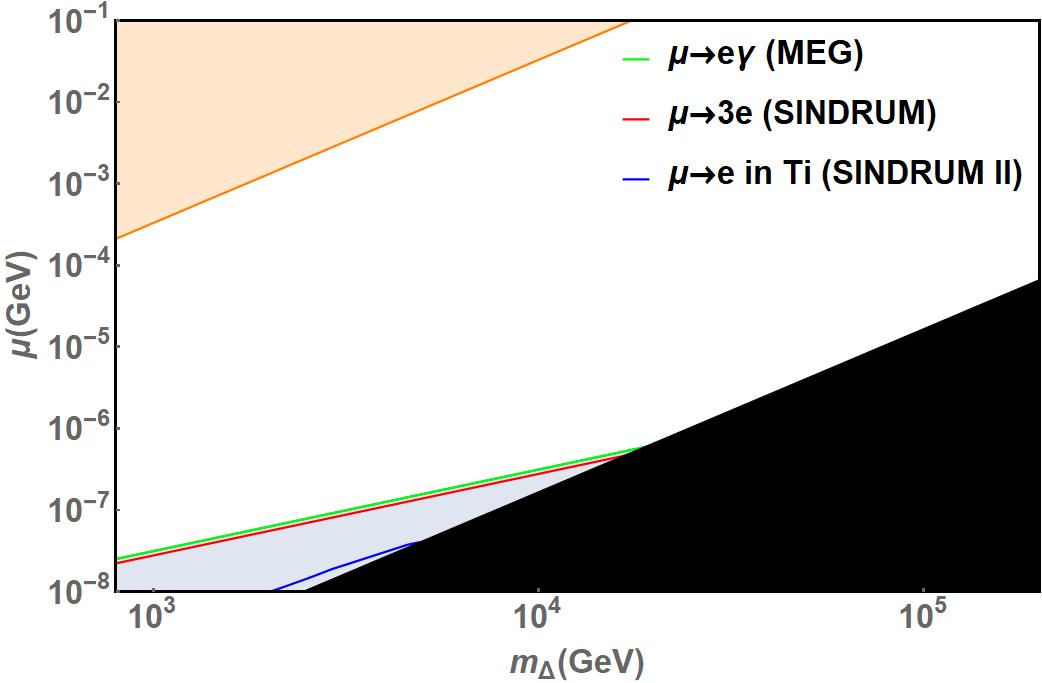}}
    \subfigure[Future Sensitivities]{\includegraphics[height=0.23\textheight,width=0.49\textwidth]{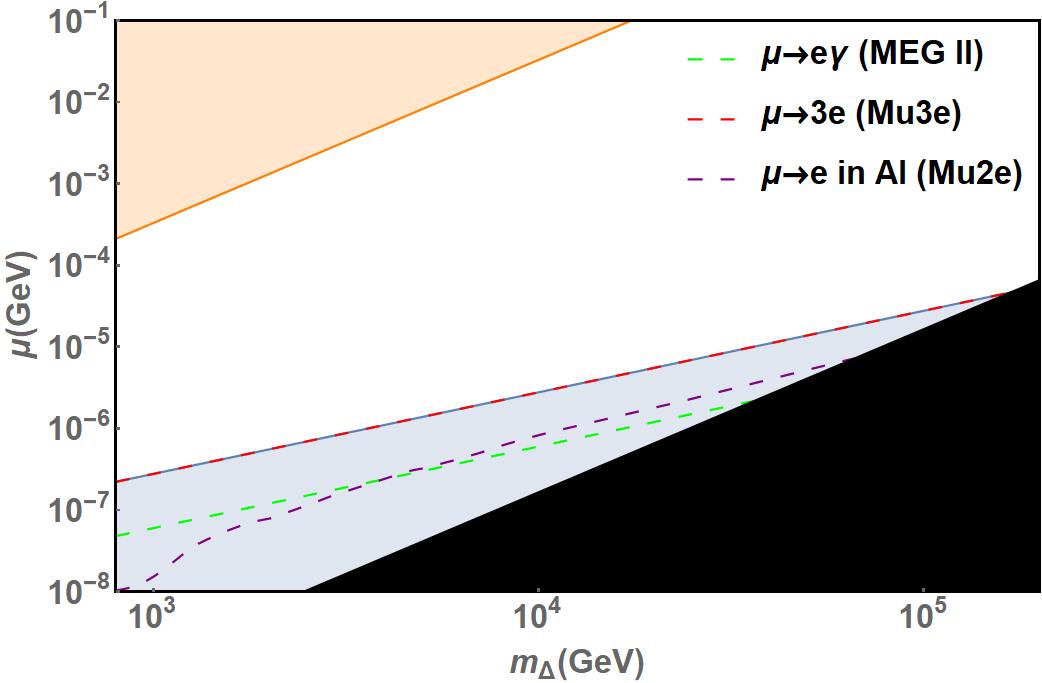}}
    \caption{\label{fig:result_io}The same as Fig.\,\ref{fig:result_no} in the IO scenario.}
\end{figure}

In this subsection we summarize our previous analyses and present the final results. The current constraints on the parameter space of $m_\Delta$ and $\mu$ from different CLFV processes are shown in Fig. \ref{fig:result_no} (a) and Fig. \ref{fig:result_io} (a) for the NO and IO scenario respectively. The constraints from requiring perturbative Yukawa couplings and avoiding the wash out effect are also depicted. As is discussed in previous subsections, the constraints are derived from scanning the parameter region which is consistent with neutrino oscillation data with Monte Carlo method and adopting the points with minimal predicted decay rates, which are translated to the most conservative constraints on the model. It is shown that in the NO scenario, the most stringent constraint comes from $\mu\rightarrow e\gamma$ decay at MEG, while in the IO scenario, the most stringent constraint is derived from $\mu\rightarrow e\gamma$ decay at MEG and $\mu\rightarrow 3e$ decay at SINDRUM, which are close to each other.

We also illustrate the sensitivities of future CLFV experiments in Fig. \ref{fig:result_no} (b) and Fig. \ref{fig:result_io} (b). In the NO scenario, $\mu\rightarrow 3e$ decay at Mu3e and $\mu\rightarrow e$ conversion in Al at Mu2e have comparable sensitivities, which are more sensitive than $\mu\rightarrow e\gamma$ at MEG II. In the IO scenario, the most sensitive experiment will be Mu3e.

\section{Discussion and Conclusion}
\label{conclusion}
In this paper, we focus on the charged lepton flavor violation (CLFV) tests within the framework of type II seesaw leptogenesis. The key processes considered include $\mu^+\to e^+\gamma$, $\mu^+ \to e^+e^-e^+$ and $\mu^- N \to e^- N$. We performed a comprehensive parameter space scan using the Monte Carlo method and derived conservative constraints on the parameters of type II seesaw leptogenesis that are consistent with current neutrino oscillation data. Our results indicate that, in the normal ordering (NO) scenario, the most stringent constraint arises from the $\mu\rightarrow e\gamma$ as measured by the MEG experiment. In the inverted ordering (IO) scenario, the strongest constraints are derived from $\mu\rightarrow e\gamma$ decay at MEG and $\mu\rightarrow 3e$ decay at SINDRUM, with both providing similar sensitivities.
We also analyzed the potential sensitivities of upcoming CLFV experiments. For the NO scenario, $\mu\rightarrow 3e$ decay at Mu3e and $\mu\rightarrow e$ conversion in aluminum nuclei at Mu2e are predicted to have comparable sensitivities, both surpassing the sensitivity of $\mu\rightarrow e\gamma$ at MEG II. In the IO scenario, the most sensitive experiment is expected to be Mu3e.

\addcontentsline{toc}{section}{Acknowledgments}
\section*{Acknowledgements}
C.\,H.\ acknowledges supports from the National Key R{\&}D Program of China under grant 2023YFA1606100, the Sun Yat-Sen University Science Foundation, the Fundamental Research Funds for the Central Universities at Sun Yat-sen University under Grant No.\,24qnpy117, 
 and the Key Laboratory of Particle Astrophysics and Cosmology (MOE) 
of Shanghai Jiao Tong University.

\addcontentsline{toc}{section}{References}
\bibliographystyle{JHEP}
\bibliography{reference}
\end{document}